\documentclass[twocolumn]{aastex701}
\usepackage{color}
\usepackage[titletoc]{appendix}
\usepackage{amsmath}
\usepackage{amssymb}
\usepackage{mathtools}
\usepackage{upgreek}
\usepackage{float}
\usepackage{comment}
\usepackage{enumitem}
\usepackage{natbib}
\usepackage{graphicx}
\usepackage{bm}
\usepackage{totcount}
\usepackage{multirow}
\usepackage{hyperref}
\usepackage{cleveref}
\usepackage{tikz}
\usepackage{soul}
\usepackage{xcolor}

\newtotcounter{citnum} 
\def\oldbibitem{} \let\oldbibitem=\bibitem
\def\bibitem{\stepcounter{citnum}\oldbibitem}

\shortauthors{Suar \& Millholland}
\shorttitle{Tilting Planets from Stellar Oblateness}

\begin{document} 

\title{Planetary Obliquity Excitation Through Pre-Main Sequence Stellar Evolution}

\author[0009-0000-3694-8993]{Sidhant Kumar Suar}
\affiliation{Department of Physics and Astronomy, University College London, London, WC1E 6BT, UK}
\affiliation{David A. Dunlap Department of Astronomy and Astrophysics, University of Toronto, Toronto, ON M5S 3H4, Canada}
\email{sidhant.suar.24@ucl.ac.uk}

\author[0000-0003-3130-2282]{Sarah C. Millholland}
\affiliation{Department of Physics, Massachusetts Institute of Technology, Cambridge, MA 02139, USA}
\affiliation{MIT Kavli Institute for Astrophysics and Space Research, Massachusetts Institute of Technology, Cambridge, MA 02139, USA}
\email{sarah.millholland@mit.edu}

\begin{abstract}
A planet's axial tilt (``obliquity'') substantially affects its atmosphere and habitability. It is thus essential to comprehend the various mechanisms that can excite planetary obliquities, particularly at the primordial stage. Here, we explore planetary obliquity excitation induced by the early evolution of the host star. A young, distended star spins rapidly, resulting in a large gravitational quadrupole moment that induces nodal recession of the planet's orbit. As the star contracts and spins down, the nodal recession frequency decreases and can cross the planet's spin axis precession frequency. An adiabatic encounter results in the planet's capture into a secular spin-orbit resonance and excites the obliquity to large values. We find planets within $a \lesssim 1 \ \mathrm{AU}$ are most affected, but adiabatic capture depends on the initial stellar radius and spin rate. The overall picture is complicated by other sources of perturbation, including the disk, multiple planets, and tidal dissipation. Tides make it such that stellar oblateness-induced obliquity excitation is transient since tidal perturbations cause the resonance to break once high obliquities are reached. However, this early transient excitation is important because it can prime planets for long-term capture in a secular spin-orbit resonance induced by planet-planet interactions. Thus, although stellar oblateness-induced resonances are short-lived, they facilitate the prevalence of long-lived non-zero obliquities in exoplanets.
\\[5pt]
\end{abstract}

\section{Introduction}
\label{sec: Introduction}


The dynamics of a planet's spin vector influence many of its properties, including atmospheric circulation, tidal evolution, and potential habitability \citep[e.g.][]{Atobe2004Icarus, Spiegel2009ApJ, Heller2011A&A, Jernigan2023ApJ}. Spin dynamics pertain to a planet's rotation rate and obliquity, defined as the angle between the planet's spin and orbital axes. Exoplanet obliquities are not yet widely observable, although some progress has been made. A few directly-imaged planets or brown dwarfs now have constraints on their obliquities through combining photometric, spectroscopic, and imaging data \citep{Bryan2020AJ, Bryan2021AJ, 2024AJ....168..270P}. For transiting planets, the obliquity may be constrained by measuring the deviation in the transit light curve due to the planet's rotationally-induced oblateness \citep{Seager2002ApJ, Barnes2003ApJ, Carter2010ApJ, Biersteker2017AJ, Akinsanmi2020MNRAS}. Recent studies, including some using JWST data, have used this method to place degenerate limits on the oblateness parameter, obliquity, and spin rate \citep{2024ApJ...977L...1L, 2024ApJ...976L..14L, 2025ApJ...981L...7P}, but more precise constraints require further data.

While we await more observational constraints on obliquities, we rely on theory to probe spin dynamics and make predictions that future observations can test. In this paper, we focus on obliquities. Planetary obliquities can be excited by several known mechanisms, almost all of which were originally discussed in the context of the Solar System \citep[e.g.][]{Tremaine1991Icarus, Ward2004AJ, Lee2007Icarus}. Recent studies have begun extending them to exoplanets. Giant impacts and mergers between proto-planets can produce large obliquities early on during the system's evolution \citep{Li2020ApJ}. For example, it is thought that the $98^{\circ}$ tilt of Uranus was a result of one or more giant impacts during its formation \citep{Korycansky1990Icarus, Morbidelli2012Icarus, Kegerreis2018ApJ}. Other mechanisms include planet-planet scattering \citep{Li2021ApJ, Hong2021ApJ}, disk fragmentation \citep{Jennings2021MNRAS}, and interactions with tilted circumplanetary disks \citep{Martin2021ApJ}.

In this work, we focus on a widely-studied mechanism called secular spin-orbit resonance, in which the frequencies of a planet's orbit nodal precession and its spin axis precession evolve and become commensurable, resulting in potentially significant obliquity excitation \citep{Laskar1993Nature, Ward2004AJ, Hamilton2004AJ, Correia2015A&A, Millholland2019Nature, Su2020AJ, Yuan2024arXiv}. Secular spin-orbit resonances may be encountered through various mechanisms that change the orbital and/or spin axis precession frequencies over time. One such mechanism is orbital migration; for instance, it has been shown that planets migrating into mean-motion resonances can also cross secular spin-orbit resonances along the way \citep{Brasser2015AJ,
Vokrouhlicky2015ApJ, Millholland2019Nature, Millholland2024ApJ}. Another mechanism is tidal evolution. Short-period planets have a fast tidal damping timescale of the spin axis and can evolve directly into the resonance \citep{Ward1975AJ, Millholland2020ApJ, Su2022MNRAS}. At longer periods, the tidal evolution of a planetary moon can change the spin axis precession frequency and excite the planet into the resonance \citep{Wisdom2022AAAS, Saillenfest2023}. Finally, secular spin-orbit resonance can also occur when the planet is still embedded in the protoplanetary disk. The disk induces the planet's orbital precession at a frequency that decreases as the disk dissipates, and if it becomes commensurable with the spin axis precession frequency, it can capture into resonance \citep{Millholland2019ApJ, Su2020AJ}.

The disk dissipation just described is not the only influence that can affect planets early on in the system's lifetime. Any other external perturbation that causes a deviation from the $1/r$ central potential will also lead to orbit nodal recession. A pre-main sequence (PMS) star has a large radius and rapid rotation \citep{Shu1987ARAA}, which creates a significant gravitational quadrupole potential and drives regression of the planet's longitude of ascending node \citep{Murray1999ssd}. The precession frequency decreases as the star spins down and contracts \citep{Bouvier2014prpl}. The consequences of this process have been studied in a variety of extrasolar contexts, in particular as a mechanism to tilt orbits \citep{Batygin2013ApJ, Spalding2016ApJ, Spalding2017AJ, LiG2020ApJ, Spalding2020AJ,
Becker2020AJ, Schultz2021MNRAS, Chen2022ApJ, Faridani2023ApJ, 2025ApJ...978...18F}. However, a comprehensive study of the role of the evolving stellar oblateness in planetary spin dynamics, such as which planets are most susceptible and how it interacts with other sources of perturbation, has not yet been conducted.

In this study, we focus on a young star's quadrupole potential as a generator of secular spin-orbit resonance. As the orbital precession frequency decreases and approaches the spin axis precession frequency adiabatically, it can lead to large excitations of the planet's obliquity. Therefore, the evolution of PMS stellar parameters will be important in understanding the resulting planetary spin motion. We will show that the mechanism is only relevant for close-in planets. The study of secular spin-orbit resonances from stellar oblateness is not altogether new; \cite{Winn2005ApJ} and \cite{Fabrycky2007ApJ} discussed this idea in the context of exploring whether obliquity tides could inflate hot Jupiters, and \cite{Millholland2019Nature} explored it briefly within a broader study of secular spin-orbit resonances among Kepler multi-planet systems. In this work, we comprehensively examine the required conditions and resulting consequences of secular spin-orbit resonances from stellar oblateness, and we explore this in the context of the other factors affecting a planet's early evolution, including the disk, other planets, and tides.   

The outline of the paper is as follows. In Section \ref{sec: Spin Dynamics}, we review the evolution of pre-main sequence stars, stellar oblateness-induced nodal precession, and spin axis precession. In Section \ref{sec: Resonance}, we outline the dynamics of secular spin-orbit resonances and compute the required parameter space limits for resonance crossing and capture. We use a perturbative Hamiltonian in Section \ref{sec: Obliquity Evolution} to model the secular evolution of planetary obliquities over a wide range of parameters to verify these limits. In Section \ref{sec: N-body}, we use an \textit{N}-body code to verify our secular evolution results and evolve more complicated simulations including perturbations from the disk, multiple planets, and tides. Finally, we discuss the results and conclude in Section \ref{sec: Conclusion}.


\section{Stellar Evolution and Spin Dynamics}
\label{sec: Spin Dynamics}
A young star's rapid rotation leads to an equatorial bulge, causing the star to become an oblate spheroid. This deformation introduces a quadrupole component to the star's gravitational potential. The stellar quadrupole moment ($J_2$) drives the regression of the planet's longitude of ascending node. It is thus crucial to understand the time evolution of the stellar oblateness, which depends on the evolution of its spin rate and radius. In this section, we construct models for the host star's evolution and define the key frequencies relevant to the planet's orbital and spin precession. 

\subsection{Stellar Evolution}
Our stellar evolution model is primarily drawn from \cite{Batygin2013ApJ}, who modeled the pre-main sequence rotational evolution by accounting for the
competing effects of spin-up due to gravitational contraction and spin-down due to magnetic star-disk coupling. The details of the model are presented in Appendices \ref{sec: Kelvin-Helmholtz Contraction} $\&$ \ref{sec: Angular Momentum Transport}, but here we outline a few important aspects. 

The star is considered to be fully convective with polytropic index $n=3/2$, closely resembling a PMS solar-like star. In the later phases of stellar evolution, a star might develop a radiative core, but those timescales are not as relevant as the ones we are concerned with in this paper. The star undergoes Kelvin-Helmholtz contraction, which shrinks its radius and causes spin-up due to angular momentum conservation. However, this is not the only rotational influence. Another factor is the rotating accretion disk around the host star, which plays a major role through star-disk angular momentum transfer. The accretion disk mass $M_{d}$ is set to evolve as a simple power law, where the disk decays with time according to
\begin{equation}
    \label{eq: disk_mass}
    M_{d} = \frac{M_{d_{0}}}{1+t/\tau_{d}}.
\end{equation} 
Here, $M_{d_{0}}$ is the initial disk mass and $\tau_{d}$ is the disk decay timescale. Although this form of disk mass decay is a time-averaged representation, it still agrees well with observations \citep{Hartmann1998ApJ, Hillenbrand2008PhST, Batygin2013ApJ}. 

The presence of the disk complicates the star's rotational evolution. Disk accretion causes spin-up, while magnetic braking from star-disk coupling causes spin-down. Additional effects like stellar and disk winds can change the rotation rate of stars, but these effects can be ignored during the disk-bearing phase, especially under disk-locked conditions in which there is synchronized rotation between the star and the truncation radius of the disk, and where we assume that the accretion disk gains all the angular momentum lost by the protostar. Winds either carry away a small fraction of the disk mass or have smaller torques \citep{Matt2008ApJ} compared to other mechanisms during the disk-bearing phase.
\begin{figure}[ht!]
\centering
    \includegraphics[width=0.48\textwidth]{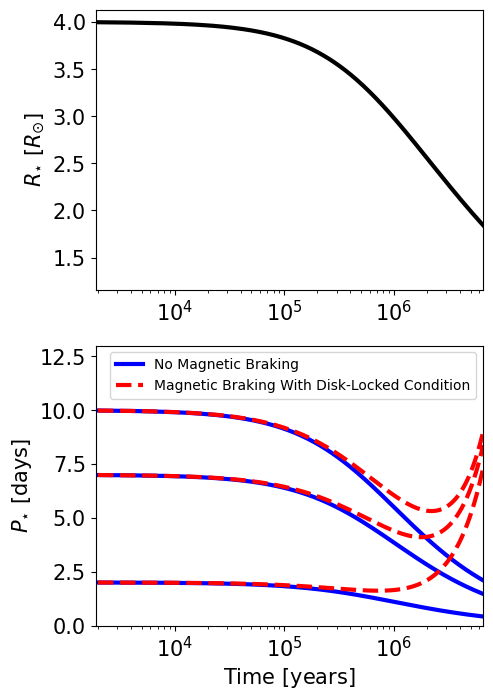}
    \caption{\textit{Top panel}: Evolution of the stellar radius of a solar mass star ($n=3/2$ polytrope) with initial radius $4 \ R_{\odot}$ and effective temperature $4270 \ \mathrm{K}$. \textit{Bottom panel}: Corresponding rotational evolution accounting for stellar contraction and star-disk interactions. We present three initial conditions corresponding to fast, medium, and slow rotators. The magnitude of the magnetic braking dictates the stellar spin distribution in the later stages of the system's evolution, where a weak braking would lead to a spin-up due to gravitational contraction and accretion, while a more effective magnetic braking would lead to stellar spin-down.}
    \label{fig: P_rot}
\end{figure}

Figure \ref{fig: P_rot} shows the evolution of a pre-main sequence star's radius and rotation period. Gravitational contraction dominates the rotational evolution in the early stages, while magnetic braking in disk-locked conditions takes over during the later stages. As the star contracts with time, the ratio of the hydrostatic pressure in the stellar interior to the magnetic pressure at the stellar surface decreases, leading to more efficient magnetic braking.

The magnetic torque slows the star down, and its rotation period agrees well with observed values \citep[e.g.][]{Affer2013MNRAS}. It is important to note that these calculations of stellar evolution involve several approximations, but they agree relatively well with observed PMS stellar rotation distributions and are sufficiently accurate for the purposes of our study of stellar oblateness-induced secular spin-orbit resonance.

\subsection{Orbit Nodal Recession}
The quadrupole moment of the rapidly rotating young star induces nodal recession of the planet's orbit with period $T_{g} \equiv 2\pi/g$. For a planet in circular orbit with semi-major axis $a_{p}$, the stellar-induced nodal regression rate due to a PMS star with spin rate $\omega_{\star}$, radius $R_{\star}$, and Love number $k_{2_{\star}}$ is given by \citep{Murray1999ssd}
\begin{equation}
    \label{eq: g}
    g = -\dot{\Omega} = n_{p}\frac{k_{2_{\star}}}{2}\left(\frac{\omega_{\star}}{n_{p}}\right)^{2}\left(\frac{R_{\star}}{a_p}\right)^{5},
\end{equation}
where $n_{p} = \sqrt{GM_{\star}/a_{p}^{3}}$ is the mean motion and $\Omega$ is the longitude of the ascending node with its time derivative $\dot{\Omega} < 0$. This definition for $g$ already accounts for the star's gravitational quadrupole moment \citep{Sterne1939MNRAS, Spalding2016ApJ}
\begin{equation}
    J_{2_{\star}} = \frac{1}{3}\left(\frac{\omega_{\star}}{\omega_{b_{\star}}}\right)^{2}k_{2_{\star}},
\end{equation}
where $\omega_{b_{\star}} = \sqrt{GM_{\star}/R_{\star}^{3}}$ is the star's break-up angular velocity.

\subsection{Spin Axis Precession}
Alongside the planet's orbital precession, torques from the star also cause the planet's spin angular momentum vector to precess about its drifting orbital angular momentum vector. The precessional period is
\begin{equation}
    T_{\alpha} \equiv \frac{2\pi}{\alpha|\cos{\epsilon}|}
\label{eq: T_alpha}
\end{equation}
where $\epsilon$ is the planet's obliquity and $\alpha$ is the spin axis precession constant. In the absence of satellites and a circumplanetary disk, $\alpha$ is given by \citep{Millholland2019ApJ, Su2020AJ, Su2022MNRAS}
\begin{equation}
    \alpha = \frac{1}{2}\left(\frac{n_{p}^{2}}{\omega_p}\right)\left(\frac{k_{2p}}{I_{p}}\right)\left(\frac{\omega_p}{\omega_{bp}}\right)^{2}.
    \label{eq: alpha}
\end{equation}
Here $\omega_p$ is the planet's spin rate, and $\omega_{bp} = \sqrt{GM_{p}/R_{p}^{3}}$ is its break-up angular velocity. The quantity $k_{2p}$ is the planet's Love number, a dimensionless quantity that measures the planet's deformation response to external tidal potentials. $I_{p}$ is the planet's moment of inertia normalized by $M_{p}R_{p}^{2}$. This definition for $\alpha$ already accounts for the planet's gravitational quadrupole moment.

In the presence of satellites or a circumplanetary disk, the value of $\alpha$ would be enhanced by some factor, but we consider no such effects in this paper, as we deal with close-in planets that are less likely to have satellites or be surrounded by a circumplanetary disk. The coupling between the spin axis precession and the steady evolution of the orbit nodal recession can result in a resonance crossing that excites the planet's obliquity to as high as $90^{\circ}$. This requires certain conditions to be met, which we will analyze in the next section.

\section{Secular Spin-Orbit Resonance}
\label{sec: Resonance}

Secular spin-orbit resonance requires that the star-planet system crosses $T_{\alpha}/T_{g} = 1$ from above. We call this the ``crossing criterion''. As the PMS star undergoes gravitational contraction, the orbital precession rate (equation \ref{eq: g}) slows, so $T_{g}$ increases. Meanwhile, $T_{\alpha}$ is independent of time, assuming (for now) that there is no tidal evolution or disk migration. This implies that crossing through unity from above is inevitable if the system was at $T_{\alpha}>T_{g}$ during the early phases of its evolution. 

If the planet begins with an initial obliquity close to $0^{\circ}$, the crossing criterion can be reduced to ${(g/{\alpha})_0\geq1}$ (where the subscript indicates the initial condition). Assuming a set of fiducial parameters indicated in Table \ref{table: Model Parameters} and $\omega_p = 0.1\omega_{bp}$, we use equations \ref{eq: g} and \ref{eq: alpha} to simplify $g/\alpha$ to
\begin{equation}
    \label{eq: g_over_alpha}
    \frac{g}{\alpha} \approx 10\left(\frac{\omega_{\star}}{\mathrm{2\pi/day}}\right)^{2}\left(\frac{R_{\star}}{4 \ R_{\odot}}\right)^{5}\left(\frac{a_p}{0.1 \ \mathrm{AU}}\right)^{-1/2}.
\end{equation}
Setting ${(g/{\alpha})_0\geq1}$ gives the maximum semi-major axis at which the resonance will be crossed, assuming a given initial stellar spin rate $\omega_{\star 0}$ and radius $R_{\star 0}$. This yields
\begin{equation}
    \label{eq: a_max_crossing}
    a_p \lesssim 9.86 \ \mathrm{AU}\left(\frac{\omega_{\star 0}}{\mathrm{2\pi/day}}\right)^{4}\left(\frac{R_{\star 0}}{4 \ R_{\odot}}\right)^{10}
\end{equation}
for resonance crossing.

Resonance crossing is a necessary, but not sufficient, condition for resonance capture. Successful resonance capture also requires that the crossing occurs adiabatically, meaning the timescale for crossing must be longer than the resonant libration period of the planet's spin axis \citep{Hamilton2004AJ, Tremaine2023dyps}. We call this the ``capture criterion''. The conditions for both resonance crossing and capture are best expressed through a set of inequalities,
\begin{equation}
    \label{eq: adiabatic}
    \mathrlap{\overbrace{\phantom{0 < \dot{\alpha}-\dot{g}}}^{\text{crossing}}}
    0 < 
    \mathrlap{\underbrace{\phantom{\dot{\alpha}-\dot{g} \lesssim \alpha g\sin{\epsilon_{\mathrm{cr}}}\sin{i}}}_{\text{capture}}}
    \dot{\alpha}-\dot{g} \lesssim
    \alpha g\sin{\epsilon_{\mathrm{cr}}}\sin{i},
\end{equation}
where $\epsilon_{\mathrm{cr}}$ is the planet's obliquity at resonance crossing and $i$ is the orbital inclination. Resonance capture requires a non-zero inclination and obliquity at the resonance crossing. 

\begin{table}[t!]
    \centering
    \begin{tabular}{ccc}
        \multicolumn{3}{c}{\textbf{Table 1}} \\
        \multicolumn{3}{l}{Fiducial Parameters Used in All Calculations} \\
        \hline
        \hline
        & Symbol &\hspace{20pt} Model \\
        \hline
        & $M_{d_{0}}$ &\hspace{20pt} $0.01 \ \mathrm{M_{\odot}}$ \\
        & $\tau_{d}$ &\hspace{20pt} $0.5 \ \mathrm{\mathrm{Myr}}$ \\
        \hline
        & $M_{\star}$ &\hspace{20pt} $1 \ \mathrm{M_{\odot}}$ \\
        & $T_{\mathrm{eff}}$ &\hspace{20pt} $4270 \ \mathrm{K}$ \\
        & $k_{2_{\star}}$ &\hspace{20pt} $0.28$ \\
        & $I_{\star}$ &\hspace{20pt} $0.2$ \\
        & $\Gamma$ &\hspace{20pt} $1$ \\
        & $\lvert\vec{B}_{\star}(R_{\star},\pi/2)\rvert$ &\hspace{20pt} $0.15 \ \mathrm{Tesla}$ \\
        & $\beta$ &\hspace{20pt} $0.01$ \\
        \hline
        & $M_{p}$ &\hspace{20pt} $10 \ \mathrm{M_{\oplus}}$ \\ 
        & $R_{p}$ &\hspace{20pt} $3 \ \mathrm{R_{\oplus}}$ \\
        & $k_{2_{p}}$ &\hspace{20pt} $0.3$ \\
        & $I_{p}$ &\hspace{20pt} $0.3$ \\
        & $i$ &\hspace{20pt} $5^{\circ}$ \\
        & $\epsilon_{\mathrm{cr}}$ &\hspace{20pt} $20^{\circ}$ \\
        & $Q$ &\hspace{20pt} $10^{4}$ \\
        \hline
    \end{tabular}
    
    \makeatletter\def\@currentlabel{1}\makeatother
    \label{table: Model Parameters}
\end{table}

We can simplify the capture criterion by assuming $\dot{\alpha} \approx 0$. Although $\alpha$ can evolve due to tidal dissipation, disk migration, and planet radius contraction, these processes often happen slower than those causing $g$ to change. Similarly, $g > 0$ while $\dot{g} < 0$ throughout the system's evolution, so we can use $-\dot{g} = \lvert\dot{g}\rvert$ which implies that $-g/\dot{g} = \lvert g/\dot{g}\rvert$ to simplify our notations. Using equation \ref{eq: g}, we can express $g/\dot{g}$ as
\begin{equation}
    \label{eq: g_over_gdot}
    \frac{g}{\dot{g}} = \left[{2\left(\frac{\dot{\omega}_{\star}}{\omega_{\star}}\right)+5\left(\frac{\dot{R}_{\star}}{R_{\star}}\right)}\right]^{-1}.
\end{equation}
The changes to the star's spin rate and radius are not straightforward to write down analytically, but if the system approaches resonance crossing adiabatically, the relevant timescales are those upon crossing. We thus use equation \ref{eq: g_over_alpha} to find the crossing time where $g/\alpha = 1$ and evaluate $(g/\dot{g})_{\mathrm{cr}}$ at that time via equation \ref{eq: g_over_gdot}. These calculations were done numerically because the relevant functions cannot be expressed analytically. The capture criterion is then rewritten as
\begin{equation}
    \label{eq: adia_simp}
    \alpha\lvert(g/\dot{g})_{\mathrm{cr}}\rvert\sin{\epsilon_{\mathrm{cr}}}\sin{i} \gtrsim 1.
\end{equation}
Finally, we can obtain the maximum semi-major axis for which the crossing is adiabatic by substituting equation \ref{eq: alpha} for $\alpha$ in equation \ref{eq: adia_simp}, resulting in
\begin{equation}
    \label{eq: a_max_capture}
    a_p \lesssim 1.35 \ \mathrm{AU}\left(\frac{\lvert(g/\dot{g})_{\text{cr}}\rvert}{1 \ \mathrm{Myr}}\right)^{1/3}.
\end{equation}

In the next calculations and those throughout this paper, we consider a fiducial system with parameters indicated in Table \ref{table: Model Parameters}. In addition, we initialize the planet's spin rate to $10\%$ of its break-up angular velocity and its obliquity to $1^{\circ}$. The initial stellar spin period and radius are set to $2 \ \mathrm{days}$ and $4 \ R_{\odot}$. 

\begin{figure}[t]
\centering
    \includegraphics[width=0.48\textwidth]{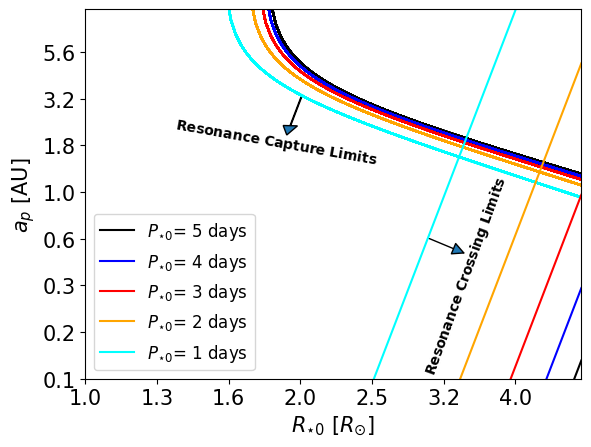}
    \caption{The limits on the planet's semi-major axis and initial stellar radius for which resonance crossing (equation \ref{eq: a_max_crossing}) and capture (equation \ref{eq: a_max_capture}) would occur. We consider multiple values of the initial stellar rotation period $P_{\star 0}$. }
    \label{fig: aR_ini_govera}
\end{figure}
Figure \ref{fig: aR_ini_govera} shows upper limits on the planet's semi-major axis needed for resonance crossing (equation \ref{eq: a_max_crossing}) and resonance capture (equation \ref{eq: a_max_capture}). We show the limits for a variety of values of the initial stellar radius $R_{\star 0}$ and the initial stellar rotation period $P_{\star 0}$. For a choice of $P_{\star 0}$, the two lines intersect to bound a region which will allow excitation to high obliquities ($\epsilon \approx 90^{\circ}$) under disk-locked conditions. The other regions are those where the system experiences either no resonance crossing (resulting in no initial excitation of the obliquity) or resonance crossing but no adiabatic capture (resulting in partial excitation). Non-adiabatic excitation occurs due to a complex phenomenon associated with the appearance of a ``separatrix'' \citep{Su2020AJ} in the level curves of the system's Hamiltonian. The degree of excitation depends on how close the crossing criterion is to the adiabatic limit. We will expand our discussion of the Hamiltonian and the corresponding evolution of obliquities in Section \ref{sec: Obliquity Evolution}.
\begin{figure}[t]
\centering
    \includegraphics[width=0.48\textwidth]{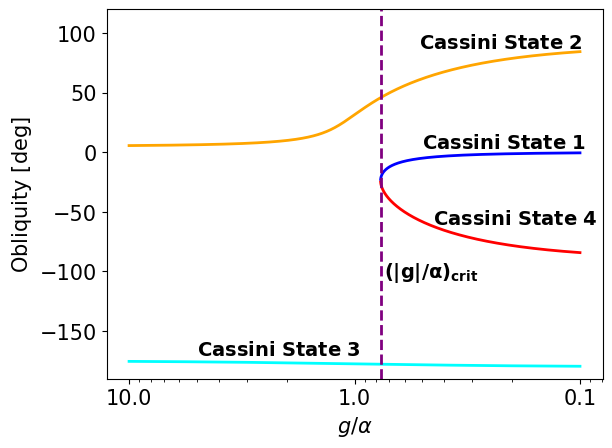}
    \caption{Obliquities as a function of $g/\alpha$ corresponding to the four different Cassini states assuming $i = 5^{\circ}$. States 1 and 4 do not exist when $g/\alpha > \left(g/\alpha\right)_{\mathrm{crit}}$, and state 2 is the only possible configuration where the planet can experience high obliquities in secular spin-orbit resonance. }
    \label{fig: Cassini_States}
\end{figure}

\subsection{\textit{Cassini States}}

Once the planet is trapped in a secular spin-orbit resonance, its spin axis is locked in an equilibrium position called a ``Cassini state'' \citep{Peale1969AJ, Ward2004AJ, Hamilton2004AJ, Tremaine2023dyps}. There are four possible Cassini states, but the resonant configuration we refer to here is Cassini state 2, in which the planet's spin and orbital angular momentum vectors precess at the same rate on opposite sides of the total angular momentum vector. We have so far assumed that there is no tidal dissipation, in which case these three vectors lie on the same plane. The Cassini states satisfy the relation \citep{Ward2004AJ, Su2020AJ},
\begin{equation}
    \label{eq: cassini} \alpha\cos{\epsilon}\sin{\epsilon}+g\sin{(\epsilon-i)} = 0,
\end{equation}
assuming that the planet’s orbit is undergoing uniform precession. Even when that's not the case, this classical approximation still holds well \citep{Fabrycky2007ApJ, Correia2015A&A, Su2020AJ}. Equation \ref{eq: cassini} has two roots if $g/\alpha > (g/\alpha)_{\mathrm{crit}}$ or four roots if $g/\alpha < (g/\alpha)_{\mathrm{crit}}$, where the critical value of $g/\alpha$ is equal to \citep{Ward2004AJ}
\begin{equation}
    \left(\frac{g}{\alpha}\right)_{\mathrm{crit}} = \left[\sin^{\frac{2}{3}}{i}+\cos^{\frac{2}{3}}{i}\right]^{-3/2}.
\end{equation}
The four roots correspond to Cassini states 1 to 4, while the two correspond to Cassini states 2 and 3. Figure \ref{fig: Cassini_States} shows the Cassini states as a function of $g/\alpha$ for ${i=5^{\circ}}$. The Cassini state 2 curve shows the path by which a planet attains a high obliquity state, starting with $g/\alpha \gg 1$ and increasing as $g/\alpha$ decreases and crosses through unity.

\section{Secular obliquity evolution}
\label{sec: Obliquity Evolution}
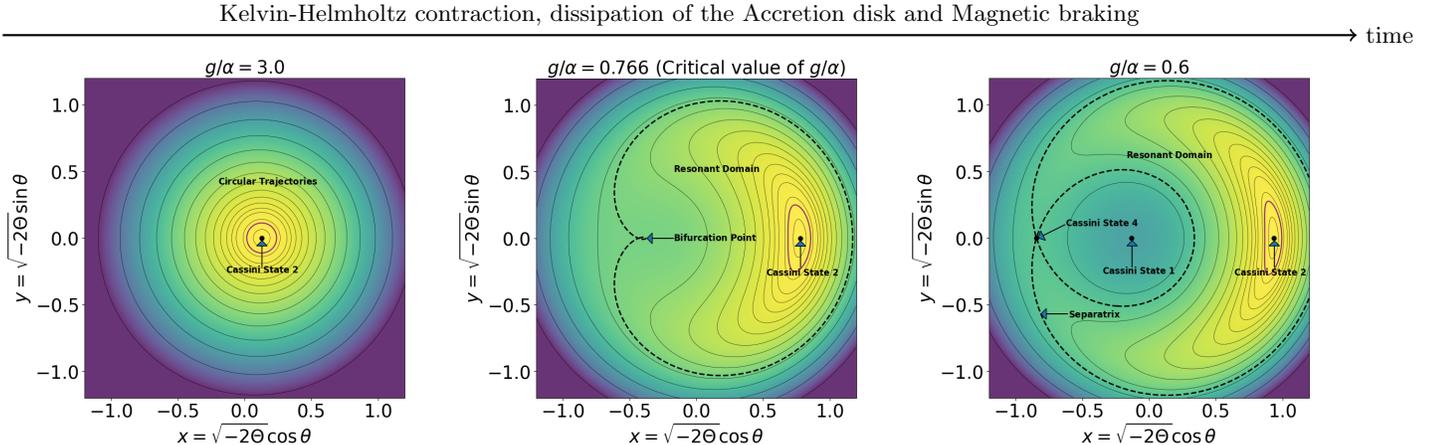
\begin{figure*}
  \begin{tikzpicture}
    \draw[thick, ->] (0,0) -- (18,0) node [near start, midway, above] {Kelvin-Helmholtz contraction, dissipation of the Accretion disk and Magnetic braking} node [right] {time};
  \end{tikzpicture}
  \begin{minipage}[t]{0.3\linewidth}
    \includegraphics[width=\linewidth]{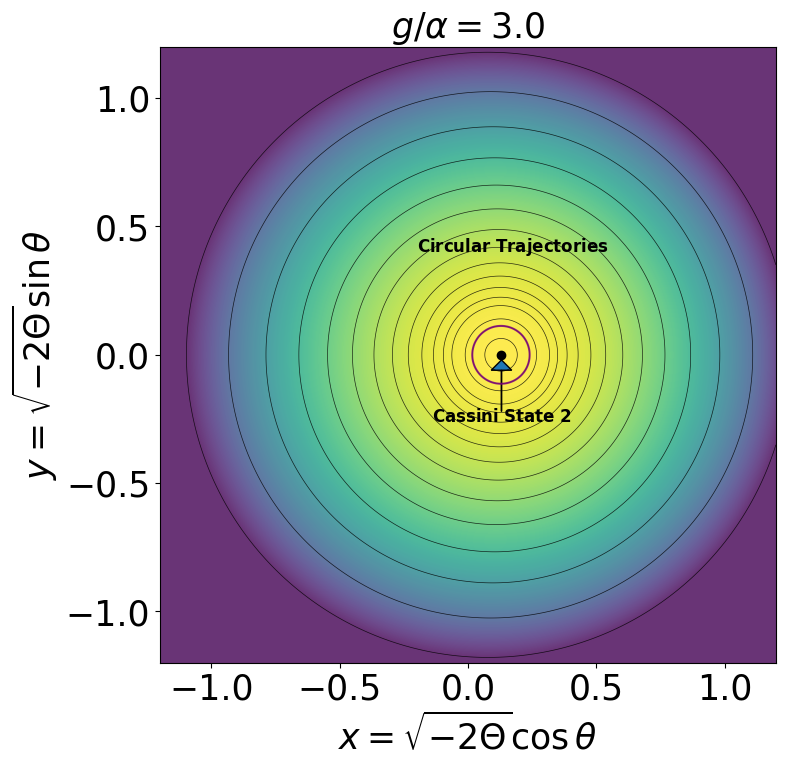}
  \end{minipage}\hfil
  \begin{minipage}[t]{0.3\linewidth}
    \includegraphics[width=\linewidth]{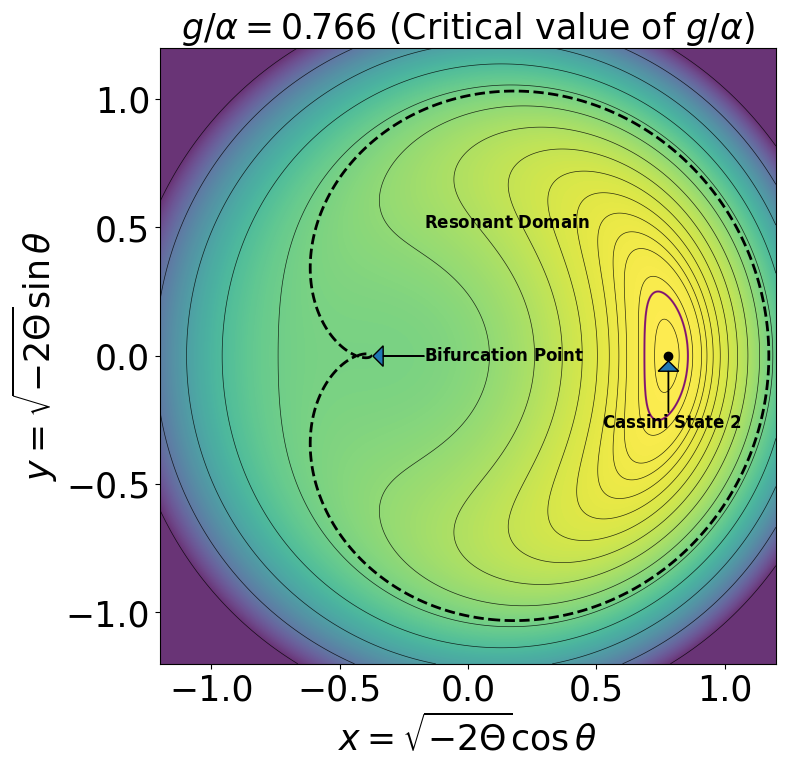}
  \end{minipage}\hfil
  \begin{minipage}[t]{0.3\linewidth}
    \includegraphics[width=\linewidth]{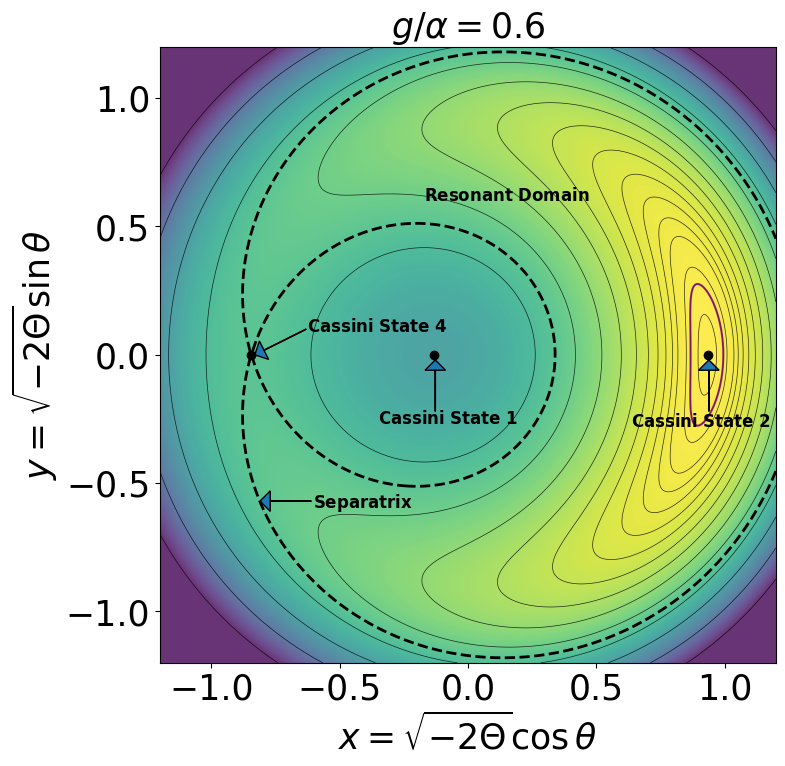}
  \end{minipage}
  \caption{Phase-space representation of the Hamiltonian (equation \ref{eq: H}) as the system evolves. The contour lines trace out level curves of the Hamiltonian. As time advances, the star contracts and spins down due to magnetic braking, and the disk loses its mass. This results in $g/\alpha$ decreasing and gives rise to the secular spin-orbit resonance. These plots use an inclination $i = 5^{\circ}$, the angle between the orbital and disk planes. As the system approaches the critical value of $\left(g/\alpha\right)_{\mathrm{crit}} = 0.766$, a separatrix appears, as do the various Cassini states. The radius of the phase plot increases with time, which indicates that the obliquity increases as the polar radius is equal to $\sqrt{-2\Theta}$, where $\Theta = \cos{\epsilon} - 1$.}
  \label{fig: phase_space}
\end{figure*}
The perturbative Hamiltonian that governs the dynamics of the planetary spin vector for low orbital eccentricities can be expressed using polar coordinates as \citep{Morbidelli2002mcma, Millholland2019ApJ}
\begin{equation}
    \label{eq: H}
    \mathcal{H} = \frac{\alpha}{2}\left(1+\Theta\right)^{2}-g\Theta-ig\cos{\theta}\sqrt{1-\left(1+\Theta\right)^{2}}.
\end{equation}
Here $\Theta = \cos{\epsilon}-1$ and $\theta$ is the azimuthal angle of the spin vector. The Cartesian coordinates are $(x,y) = \left(\sqrt{-2\Theta}\cos{\theta},\sqrt{-2\Theta}\sin{\theta}\right)$. Here we use this Hamiltonian to model the resonant encounters by evolving the parameters of the star-disk-planet system. Doing so enables us to analytically predict the evolution of planetary obliquities in such systems under different conditions and configurations.

We can first obtain a qualitative understanding of the spin axis dynamics by studying the Hamiltonian level curves in phase space \citep{Saillenfest2019A&A, Huang2023AJ}. The spin vector follows the trajectories on the level curves shown in Figure \ref{fig: phase_space}. Three values of $g/\alpha$ show the possible regimes as $g/\alpha$ decays due to the star's gravitational contraction and rotational evolution. Note that the extrema of the Hamiltonian correspond to the different Cassini states, and they move for various values of $g/\alpha$ in a manner consistent with Figure \ref{fig: Cassini_States}.  When the evolution of $g/\alpha$ is slower than the libration timescale, the phase-space area enclosed by a given trajectory is conserved. The system can only enter Cassini state 2 once the separatrix appears, after which the planet's spin vector librates about one of the Hamiltonian's extrema (Figure \ref{fig: phase_space}) on a banana-shaped level curve. The Hamiltonian in equation \ref{eq: H} is only formally integrable if $g$ and $\alpha$ are constants. However, if they vary with time, we can approximate the real system dynamics provided $g/\alpha$ evolves much slower than the libration period of $\theta$.

\begin{figure}[t!]
\centering
    \includegraphics[width=0.48\textwidth]{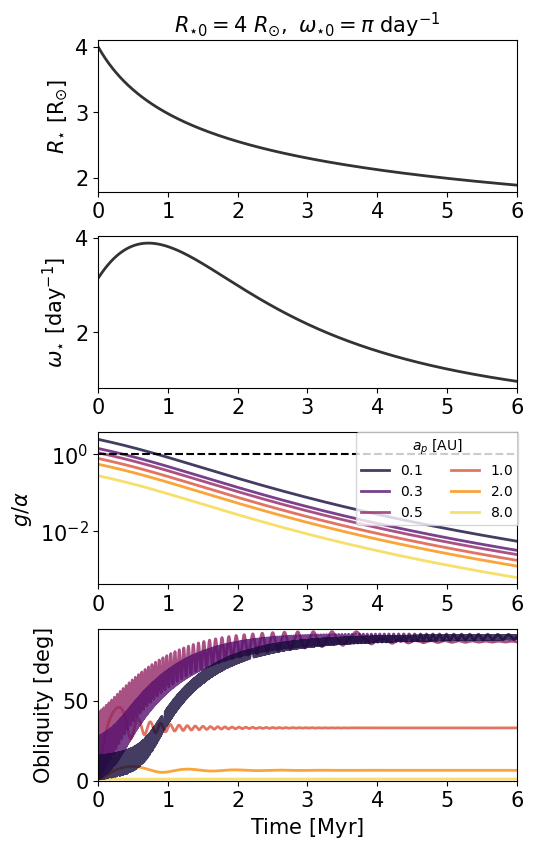}
    \caption{Planetary obliquity evolution from the secular model of the Hamiltonian in equation \ref{eq: H}. \textit{First $\&$ second panels}: Evolution of the stellar radius and spin rate of a solar-like PMS star with initial radius equal to $4 \ R_{\odot}$ and initial rotation period equal to 2 days. \textit{Third panel}: Evolution of $g/{\alpha}$, the ratio between the orbital and spin axis precession frequencies. The different colors correspond to planets with different semi-major axes. \textit{Fourth panel}: The resulting planetary obliquity evolution from the numerical integration of the Hamiltonian (equation \ref{eq: H}). These calculations use the fiducial values defined in Table \ref{table: Model Parameters}.}
    \label{fig: Obliquity_evolution_a}
\end{figure}

Here we explore the spin axis dynamics by performing numerical experiments integrating the Hamiltonian using Hamilton's equations and an explicit ODE solver. As we have seen in the previous sections, three primary parameters dictate whether the system will undergo secular resonance crossing and capture: the planet's semi-major axis, the star's initial rotation rate, and the star's initial radius.

We first simulate six cases with $a_p$ = 0.1, 0.3, 0.5, 1, 2, and 8 AU, using different values to show how the planet's obliquity evolution depends on its semi-major axis. The system parameters are the same as described earlier (after equation \ref{eq: a_max_capture}). Figure \ref{fig: Obliquity_evolution_a} shows the obliquity evolution for all cases. For $a_p =$ 1, 2, and 8 AU cases, the system does not experience resonance capture because $(g/{\alpha})_0 < 1$. The resonance capture is adiabatic for the $a_p$ = 0.1, 0.3, and 0.5 AU cases, resulting in final obliquities excited up to $90^{\circ}$. 

\begin{figure}[ht!]
\centering
    \includegraphics[width=0.48\textwidth]{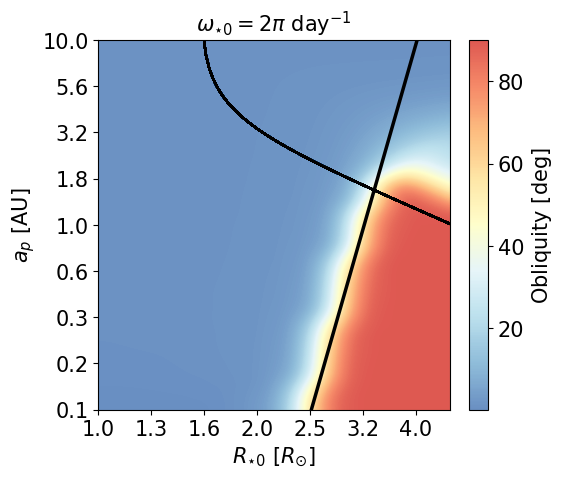}
    \caption{Map of the final planetary obliquity resulting from secular spin-orbit resonance encounters at different semi-major axes and initial stellar radii for a PMS star with an initial rotation period of $1 \ \mathrm{day}$. The color represents the final value of the obliquity that we obtain from the integration of the Hamiltonian (equation \ref{eq: H}), while the thick black lines correspond to the upper limits on the semi-major axis calculated analytically (equations \ref{eq: a_max_crossing} \& \ref{eq: a_max_capture}). The region satisfying both conditions experiences obliquity excitations up to $90^{\circ}$.}
    \label{fig: final obliquities}
\end{figure}

We can expand this investigation by running more simulations across a grid of $a_p$ and $R_{\star 0}$. We fix $\omega_{\star 0} = 2\pi \ \mathrm{day}^{-1}$ for this exercise. After running the integrations for 6 Myr, we plot the final obliquities in Figure \ref{fig: final obliquities}. The black lines correspond to the analytic crossing and capture limits derived in equations \ref{eq: a_max_crossing} \& \ref{eq: a_max_capture}. The analytic and numerical results agree reasonably well, but not perfectly. This is because, as shown in equation \ref{eq: g_over_gdot}, the relevant timescale for the decrease in the orbital precession rate is a combination of the timescales associated with the decrease in the stellar rotation rate and radius. It is thus a function of time near resonance crossing, but in our analytical approximation, we estimated it to be a constant with its value equal to $(g/\dot{g})_{\mathrm{cr}}$.

The region with $a_p$ greater than the upper limit set by the crossing criterion (equation \ref{eq: a_max_crossing}) results in no obliquity excitation. The region that obeys the former criterion but has $a_p$ greater than the upper limit set by the capture criterion (equation \ref{eq: a_max_capture}), experiences smaller excitation since the resonance crossing is too fast to be adiabatic. Finally, the region that obeys both limits results in final obliquities close to $90^{\circ}$. Exciting obliquities without resonance capture is possible, but it is not significant unless the crossing is very close to the adiabatic limit. The parameter regime where high obliquity resonances matter most corresponds to small semi-major axis ($a_p \lesssim 1$ AU), large initial stellar radius, and fast initial rotation rate, because that is when the effect of the stellar quadrupole potential is maximized.

\section{\textit{N}-body Simulations}
\label{sec: N-body}
In the previous section, we predicted the planetary obliquity evolution using a secular model with a perturbed Hamiltonian. This allowed us to see the general behavior of obliquity excitation and the parameter regimes in which it is expected. However, this approach was isolated to the impact of stellar oblateness evolution and did not include the full spectrum of dynamical effects acting during a system's early evolution. Such effects include perturbations from the protoplanetary disk, tides, and neighboring planets. While it is difficult to include all of these effects in a self-consistent secular model, we can do so in a numerical integration. In this section, we construct simulations with complexity added step-by-step to observe the impact of each effect.

Our numerical spin-orbit simulations are carried out using the \textit{N}-body code first introduced in \cite{Millholland2019Nature} and then extended in \cite{Spalding2020AJ}. The integrator includes various sources of accelerations generated in the framework of \cite{Mardling2002ApJ} and integrated with the Bulirsch-Stoer algorithm. The orbital parameters are evolved using Jacobi coordinates. In addition to standard point-mass gravitational accelerations, the code can also account for other interactions like tidal forces between the star and planets, the torques and gravitational field from a protoplanetary disk, and the quadrupole field of the star. The tidal forces are modeled using equilibrium tide theory with constant time lag \citep{Eggleton1998ApJ, Leconte2010A&A}. In addition, the spin vectors of the star and planets are all evolved self-consistently. The details of the code are provided in Appendix \ref{sec: Model for spin and orbit evolution}.

To interpret the results from the simulations, we will calculate the time evolution of the orbit nodal recession frequency and spin axis precession frequency. This is necessary to observe systems getting captured into resonance. We calculate the nodal recession frequency by performing a fast Fourier transform (FFT) within a sliding window of the time series of the x-component of the planet's orbital angular momentum vector. When there is a single source of perturbation, such as the stellar quadrupole potential, there is only one frequency component to the nodal recession. The peak frequency within each window thus indicates the precession rate of that section of time, and piecing it together gives the full-time evolution. In contrast, when we include multiple sources of precession, there are multiple frequency components, and thus we will retrieve multiple peaks from the FFT. We do not use the FFT for the spin axis precession frequency and rather just calculate it analytically using equations \ref{eq: T_alpha} and \ref{eq: alpha} with an evolving $\omega_p$ and $\epsilon$ coming directly from the simulation.

\subsection{Stellar quadrupole only}
The first type of $N$-body simulation includes only the perturbation of the stellar quadrupole potential so that we may isolate its effect and validate the results obtained using the secular model. To do this, we need to provide the code with a functional form for the stellar radius and spin based on the model we developed in Section \ref{sec: Spin Dynamics}. The stellar radius evolution has an analytic form according to equation \ref{eq: Rstar}, and we set the initial radius equal to $R_{\star0} = 4 \ \mathrm{R_{\odot}}$. However, as discussed earlier, the stellar spin rate cannot be expressed analytically when magnetic braking is taken into account. We work around this by first modeling $\omega_{\star}$ numerically using our approach from Section \ref{sec: Spin Dynamics} and then applying a curve fit to obtain an approximate analytic expression for $\omega_{\star}$ as a function of time, 
\begin{equation}
\begin{split}
    \omega_{\star} &= \omega_{\star0}\left(1+\frac{t}{\tau_{R_{\star}}}\right)^{2/3}\exp{\left(\frac{1}{2}\frac{1}{(1+\Gamma)}\frac{M_{d_{0}}}{M_{\star}}\frac{t}{t+\tau_{d}}\right)} \\
    & \left(1+\frac{c_{1}t}{\tau_{\mathrm{mag}}}\right)^{c_{2}}\exp{\left(\frac{c_{3}t}{c_{4}t+\tau_{\mathrm{mag}}}\right)},
\end{split}
\end{equation}
where we use $\omega_{\star 0} = \pi  \ \mathrm{day}^{-1}$ and the constants $c_{1} = 3.0$, $c_{2} = -2.489$, $c_{3} = 12.491$ and $c_{4} = 8.0$ are the fitting parameters. This analytical form of $\omega_{\star}$ has been inspired by equation \ref{eq: accr_omega}, which only accounts for the spin-up of the stellar rotation rate, but we modify it with two terms accounting for the damping of $\omega_{\star}$. Here, $\tau_{R_{\star}}$ is the gravitational contraction timescale, $\tau_{\mathrm{mag}}$ is the magnetic braking timescale, and $\Gamma$ is a dimensionless ratio between the components of the stellar magnetic field of a solar-type PMS star (see Appendices \ref{sec: Kelvin-Helmholtz Contraction} $\&$ \ref{sec: Angular Momentum Transport} for more details). We use the same initial stellar and planetary parameters as in Figure \ref{fig: Obliquity_evolution_a} to produce an exact comparison.
\begin{figure}[ht!]
\centering
    \includegraphics[width=0.48\textwidth]{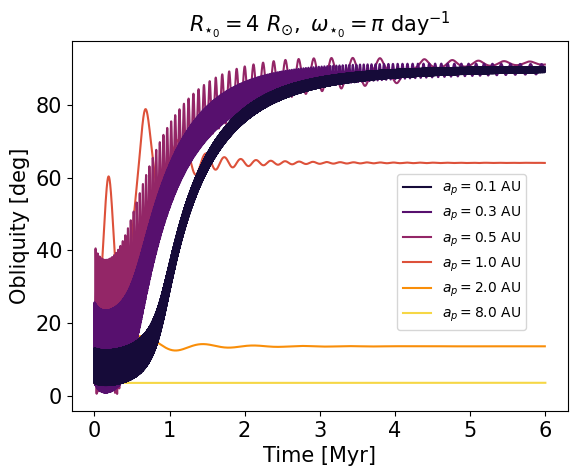}
    \caption{Planetary obliquity evolution from the $N$-body integrations that include the stellar quadrupole only. These calculations also use the same system parameters defined in Table \ref{table: Model Parameters} and thus can be directly compared to the results in Figure \ref{fig: Obliquity_evolution_a}.}
    \label{fig: Nbody_star}
\end{figure}

Figures \ref{fig: Nbody_star} and \ref{fig: Nbody_freq_star} show the results of this first type of simulation with stellar oblateness being the only source of perturbation. Figure \ref{fig: Nbody_star} includes six cases with the same semi-major axes shown in Figure \ref{fig: Obliquity_evolution_a} and thus should be directly compared. Overall, the obliquity evolution is very similar except for the borderline case $a = 1$ AU, which shows slightly more obliquity excitation in the $N$-body simulation. Figure \ref{fig: Nbody_freq_star} highlights the 0.1 AU case to show how the obliquity evolution correlates with the changes in the precession periods. Here $T_{g_{\star}}$ is the period of nodal recession where the associated frequency is calculated with the FFT, and the $\star$ subscript simply emphasizes the source of the perturbation. $T_{\alpha}$ starts greater than $T_{g_{\star}}$, but $T_{g_{\star}}$ increases due to the stellar evolution and eventually becomes equal to $T_{\alpha}$, at which point the resonance is captured and the planet's obliquity increases to $90^{\circ}$.

\subsection{Stellar quadrupole and disk}
Next, we add the gravitational potential of the disk. The early evolutionary timescales we are considering correspond to an epoch in which the disk is still present, but at a later stage. We model the disk as an infinite Mestel disk with a surface density dissipating as a power law over a timescale of $0.5 \ \mathrm{Myr}$. We adopt a uniform$-\alpha$ accretion disk model with disk surface density ($\Sigma_{0,0} = 1300 \ \mathrm{g/cm^{2}}$ at $r_{0} = 0.2 \ \mathrm{AU}$) depleted relative to the minimum-mass extrasolar nebula \citep{Chiang2013MNRAS}. The modeling details are provided in Appendix \ref{sec: Model for spin and orbit evolution}.

Figure \ref{fig: Nbody_freq_stardisk} shows the results of a simulation including the evolution of the stellar quadrupole and disk. The planet's properties are identical to those in Figure \ref{fig: Nbody_freq_star}. There are now two frequencies of orbital precession, one dominated by the star ($g_{\star}$) and one dominated by the disk ($g_d$). For the set of parameters shown here, $g_{\star}$ is orders of magnitude faster than $g_d$, and it dominates the motion. Similar to the simulation presented in Figure \ref{fig: Nbody_freq_star}, the secular spin-orbit resonance is excited when $T_{g_{\star}}$ increases and eventually becomes equal to $T_{\alpha}$. However, the behavior is different from Figure \ref{fig: Nbody_freq_star} in that the planet's spin axis eventually destabilizes when the obliquity reaches $\gtrsim70^{\circ}$, the libration amplitude increases, and the resonance is broken. This instability is a result of the multiple orbital frequency components, which become important once the planet's spin axis tilts to near $90^{\circ}$.

Although the stellar quadrupole dominates the motion in this case, more distant planets would be disk-dominated. The nodal recession frequency due to stellar oblateness decreases with the planet's semi-major axis as $g \propto a_{p}^{-7/2}$ (equation \ref{eq: g}) while the one due to the protoplanetary disk increases as $g \propto a_{p}^{3/2}$ \citep{Millholland2019ApJ}. Thus, planets on wider orbits would feel the effects of the disk and not the star.

\subsection{Stellar quadrupole, disk, and tides}
\label{subsec: Stellar quadrupole, disk, and tides}

Next, we build onto the previous simulation by now including tidal perturbations. The setup is otherwise identical, except here we model tidal effects on the planet's orbit and spin vector. The planet's tidal quality factor is set to $Q=10^4$. Figure \ref{fig: Nbody_freq_stardisk_and_tides} shows the results of this simulation. The planet gets caught in resonance with $T_{\alpha} \approx T_{g_{\star}}$ at $\sim3$ Myr and does not destabilize due to the multiple orbital frequencies, unlike the previous case. We also note that the obliquity excitation is slower than that of the simulation in Figure \ref{fig: Nbody_freq_stardisk} since the planet's spin rate $\omega_p$ slows due to tides, causing $\alpha$ to decrease as well. Provided a longer runtime, the obliquity would continue to increase, but the resonance would eventually destabilize due to overwhelming tidal torques. This phenomenon will be discussed in more detail in the next section.

\subsection{Stellar quadrupole, two planets, and tides}
\label{subsec: Stellar quadrupole, two planets, and tides}
Until now, our simulations have only considered one planet, yet most close-in planets exist in multi-planet systems. 
Here we model two planets and include perturbations from the stellar quadrupole and tides. We ignore the disk potential here, given the findings from the previous section. The inner planet is identical to that in Figures \ref{fig: Nbody_freq_star} and \ref{fig: Nbody_freq_stardisk}, and the outer planet is given the same physical properties and placed with an orbital period ratio equal to 1.9. The two planets have tidal quality factors equal to $Q = 10^4$.

\begin{figure}[ht!]
\centering
\includegraphics[width=0.48\textwidth]{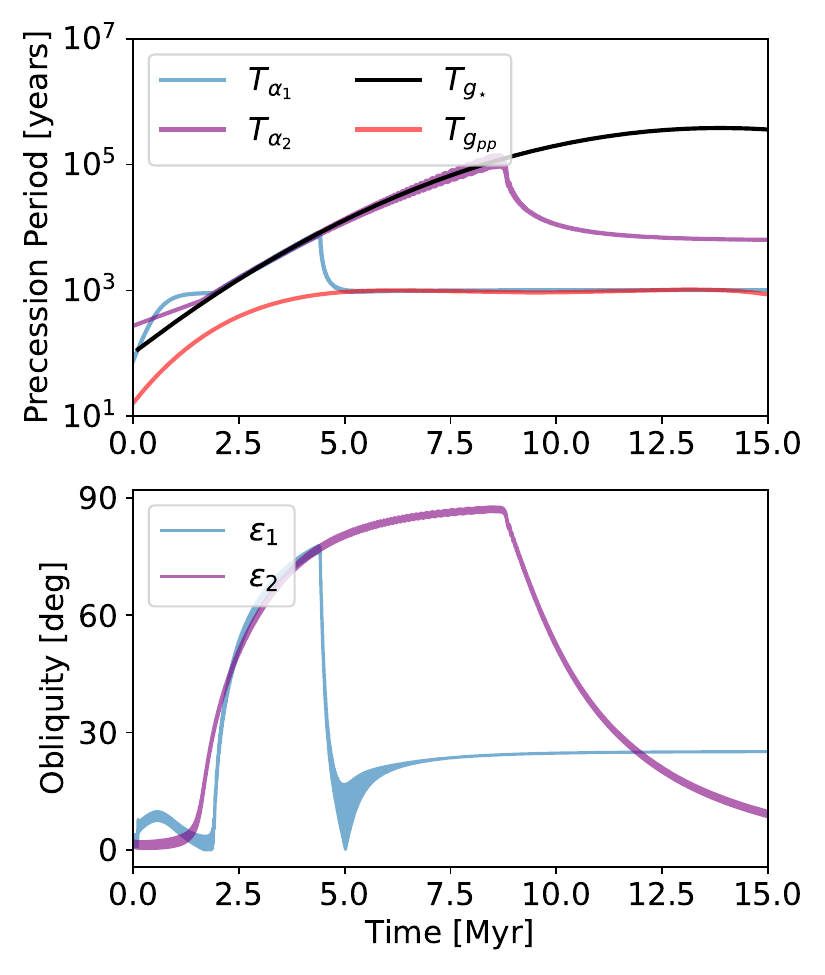}
    \caption{Results from an $N$-body integration that includes the stellar quadrupole, two planets, and tidal dissipation. The parameters of the outer planet are defined in Section \ref{subsec: Stellar quadrupole, two planets, and tides}. \textit{Top panel}: The blue and purple curves represent the spin axis precession periods of the inner and outer planets, respectively. The red and black curves represent the precession periods corresponding to the two frequency components decomposed from the planet's orbital evolution. The red curve is dominated by the planet-planet interaction, while the black curve is dominated by the star. \textit{Bottom panel}: Corresponding evolution of the obliquities of the inner and outer planets.}
    \label{fig: Nbody_freq_starplanet}
\end{figure}

Figure \ref{fig: Nbody_freq_starplanet} shows the results. There are two frequencies of orbital precession, one dominated by the star ($g_{\star}$) and one dominated by the planet-planet interactions ($g_{pp}$). Both planets get trapped into secular spin-orbit resonances with the star-dominated frequency, and their obliquities increase to large values. If tidal forces were absent in these simulations, the planets would reach and maintain $\sim90^{\circ}$, but tidal forces make it such that the obliquities cannot become arbitrarily high \citep{Fabrycky2007ApJ, Millholland2019Nature, Millholland2020ApJ, Su2022MNRAS}. The tidal torques eventually overwhelm the resonant torques keeping the planets in their Cassini states, and the resonances break at $\sim5$ Myr and $\sim9$ Myr for the inner planet and outer planet, respectively. Tides then gradually realign the spin axes, but what's most interesting is what happens after that. As the inner planet's obliquity decreases, $T_{\alpha_1}$ reaches $T_{g_{pp}}$ and the planet captures into a new resonance. The obliquity levels out at $\epsilon_1\approx25^{\circ}$. Even in the presence of tides, the planet can maintain this obliquity long-term. The outer planet's obliquity decreases but does not get captured in a new resonance, but it could in principle for different system parameters.

This simulation demonstrates an important idea: stellar oblateness-induced resonance can prime planets to get trapped in a long-term resonance controlled by the planet-planet interactions. That is, stellar radius contraction and rotational spin-down make it such that $T_{g_{\star}}$ increases so much that a planet cannot maintain an oblateness-induced resonance forever, since the planet's obliquity will eventually become too large to be tidally stable. Once the resonance destabilizes, the obliquity gradually decays and approaches the planet-planet resonance. The secular frequency dominated by the planet-planet perturbation $g_{pp}$ thus acts as a floor for the orbital precession that can yield a high obliquity resonance if a planet's spin axis precession constant $\alpha$ is larger than $g_{pp}$. This explains why the inner planet in our simulation encounters resonance, but the outer planet does not. Although necessary, this condition is not sufficient, and $T_{\alpha}$ must also approach $T_{g_{pp}}$ adiabatically from above. 

\subsection{Additional considerations}
Our $N$-body simulations have demonstrated that secular spin-orbit resonances driven by the stellar quadrupole potential lead to rich and complicated behaviors, yet there are many possible areas for further exploration. For instance, we focused on the case of a solar-like pre-main sequence star, assuming it to be fully convective, and we did not consider any major changes to its interior during its evolutionary phase. One could expand this investigation to different stellar types and also consider how a star's time-varying polytropic index affects the dynamics of resonance encounters. We also did not account for systems with more than two planets, which would introduce more orbital frequencies and complex resonance interactions.

Another effect that we did not consider is planetary mass evolution. Since the obliquity evolution we considered in this paper takes place in the first few Myr while the disk is still present, the planets may still be accreting mass. The effects on the resulting resonance encounter depend on the type of mass accretion. If it is accretion via giant impacts, the resonance would destabilize, and the planet's spin axis could get impulsively knocked to smaller or larger obliquities \citep{Li2020ApJ} before possibly encountering the resonance again later on as the star continues to evolve. 

\begin{figure}[t!]
\centering
\includegraphics[width=0.48\textwidth]{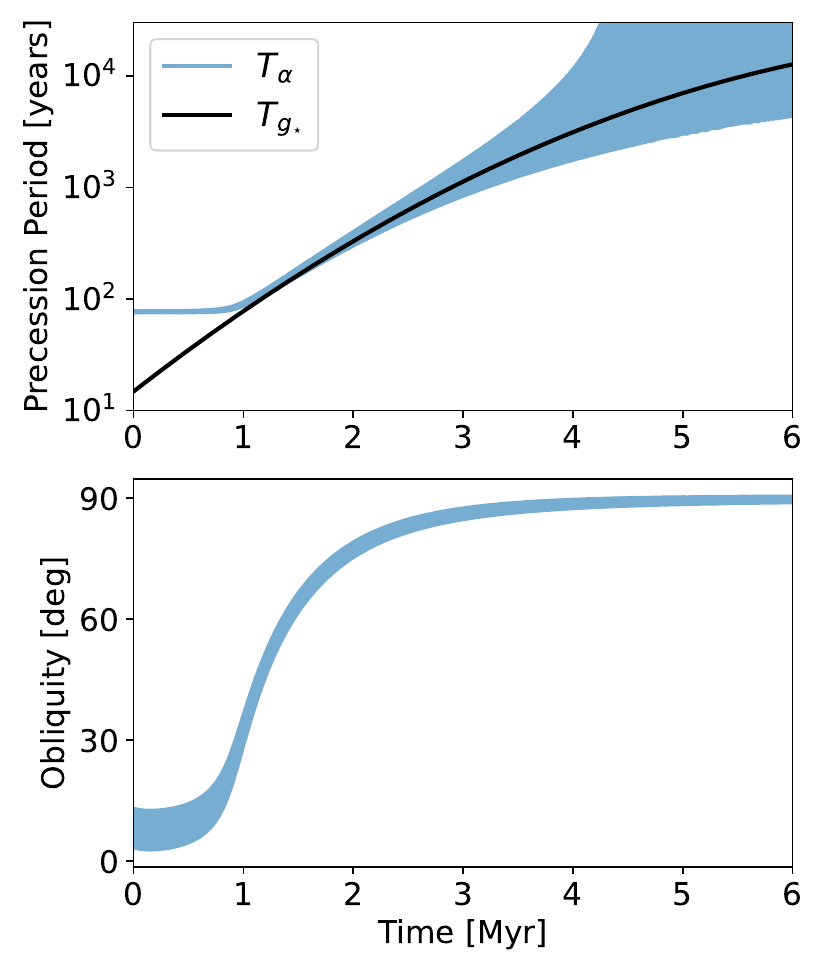}
    \caption{Results from an $N$-body integration that includes the stellar quadrupole only; this is the 0.1 AU case from Figure \ref{fig: Nbody_star}. \textit{Top panel}: Planetary spin and orbital precession periods plotted as a function of time. The blue curve is the planet's spin axis precession period, and the black curve is the nodal recession period due to the evolving PMS star's quadrupole potential. The secular spin-orbit resonance is captured when $T_{g_{\star}}$ increases and encounters $T_{\alpha}$ from below. The two stay equal as the planet is driven into the resonance. \textit{Bottom panel}: Corresponding evolution of the planetary obliquity.}
    \label{fig: Nbody_freq_star}
\end{figure}

\begin{figure}[ht!]
\centering
\includegraphics[width=0.48\textwidth]{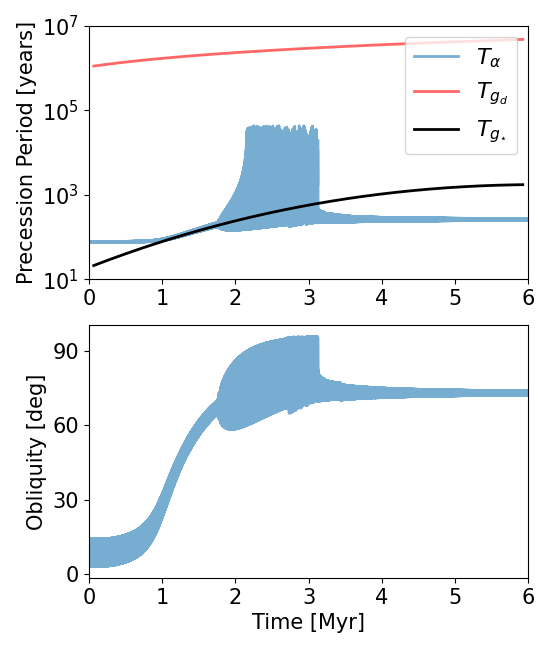}
    \caption{Results from an $N$-body integration that includes the stellar quadrupole and disk-induced precession. \textit{Top panel}: The blue curve represents the spin axis precession period of the planet. The red and black curves represent the precession periods corresponding to the two frequency components decomposed from the planet's orbital evolution. The black curve represents the contribution dominated by the star, and it is the most relevant component as it dictates the resonance capture. \textit{Bottom panel}: Corresponding evolution of the planetary obliquity, showing an increase in libration amplitude just before 2 Myr and a destabilization of the resonance just after 3 Myr.}
    \label{fig: Nbody_freq_stardisk}
\end{figure}

\begin{figure}[ht!]
\centering
\includegraphics[width=0.48\textwidth]{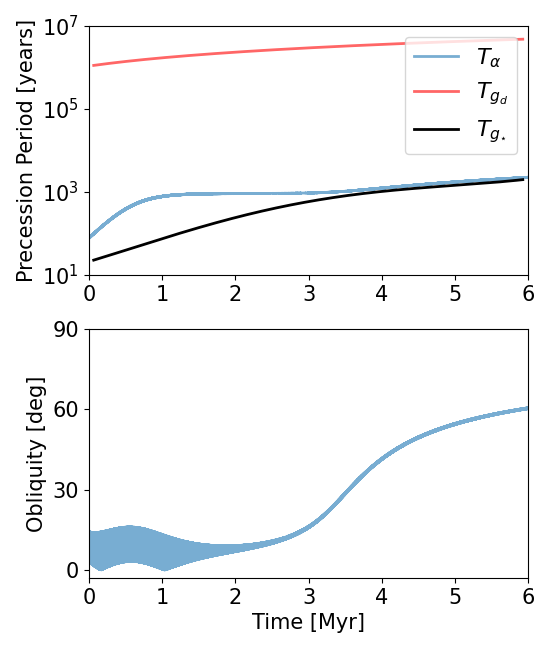}
    \caption{Results from an $N$-body integration that includes the stellar quadrupole, disk-induced precession, and tidal perturbations. Except for tides, the setup is otherwise identical to Figure \ref{fig: Nbody_freq_stardisk}. Here, the planet gets trapped in resonance around $\sim3$ Myr.}
    \label{fig: Nbody_freq_stardisk_and_tides}
\end{figure}

On the other hand, smooth gas accretion would generally be slow enough to allow the planet to maintain the resonance. Specifically, the resonance will stay stable as long as the evolution is adiabatic \citep{Henrard1982}, in which the precession timescale ($\lesssim 10^3$ yr in the first few Myr, as shown in Figures \ref{fig: Nbody_freq_star}-\ref{fig: Nbody_freq_stardisk}) is faster than the gas accretion timescale ($\sim 10^5$ yr). An increase in the planet's mass would cause $\alpha$ to decrease, such that obliquity excitation in resonance would be slower than the case with fixed planet mass. 
While beyond the scope of this paper, it would be beneficial in future work to explore a set of fully comprehensive simulations that include an evolving disk, star, and multiple planets that are undergoing mass accretion. 

\section{Conclusion}
\label{sec: Conclusion}

Secular spin-orbit resonance could be responsible for the widespread prevalence of non-zero axial tilts. Previous works focused on planet-disk interactions or planet-planet interactions as the generators of the orbital precession. Another source of perturbation is star-planet interactions early on in the system's lifetime when the star is distended and rapidly rotating. The stellar contraction and spin down lead to an evolution of the star's quadrupole moment, which can induce the capture of planets into resonance and excite their obliquities to high values. While secular spin-orbit resonances due to planet-planet interactions can last throughout the system's lifetime, those from stellar oblateness are transient, just like those from planet-disk interactions.

We find that planets with $a_p \lesssim 1 \ \mathrm{AU}$ can experience stellar oblateness-induced resonance capture, although the allowed range depends on the initial stellar radius and rotation period and specifically requires $R_{\star 0} \gtrsim 2.5 \ R_{\odot}$ (Figure \ref{fig: aR_ini_govera}). If the crossing is adiabatic and the planet does not face additional perturbations, the obliquity will reach $90^{\circ}$ since the stellar contraction and spin-down shrink the quadrupole moment to much smaller than its initial value. If the crossing is non-adiabatic, the spin axis can still tilt by a significant amount depending on the proximity to resonance capture conditions (Figures \ref{fig: Obliquity_evolution_a} and \ref{fig: final obliquities}).

In a more realistic scenario, stellar oblateness is not the only source of perturbation. We used $N$-body simulations to explore the variety of perturbations affecting early obliquity evolution, including the disk potential, stellar oblateness, multiple planets, and tidal dissipation. A case study involving just the disk potential and stellar oblateness (Figure \ref{fig: Nbody_freq_stardisk}) showed that stellar oblateness is the more dominant effect for close-in planets ($a_p \approx 0.1 \ \mathrm{AU}$). However, the disk dominates for more widely separated planets. Either way, both effects cause the planets to reach high obliquities, although the resonance can become destabilized with the competing effects of the disk potential and the stellar quadrupole potential.

In a separate case study simulation, we explored the combined effects of stellar oblateness, multiple planets, and tides (Figure \ref{fig: Nbody_freq_starplanet}). A key finding is that a resonance encounter from stellar oblateness can prime planets to subsequently enter a long-term resonance induced by planet-planet interactions. That is, stellar oblateness causes an early and dramatic tilt of a planet's spin axis to large values. Eventually, the high obliquity is destabilized due to tidal dissipation and damps back down; as it decreases, the planet encounters a different resonant commensurability where the orbital precession is driven by the planet-planet interactions, and the direction of approach is suitable for long-term capture. The early excitation thus prepares the planet for this resonance, which it may not have otherwise encountered. Therefore, even if the stellar oblateness-induced resonance cannot be sustained forever due to tides and the star not being in its PMS stage forever, these star-planet interactions are critical because they increase the probability for a planet to fall into a long-term resonance with non-zero obliquity. These findings add to a growing literature that suggests that high obliquities are prevalent in extrasolar systems.

\section{Acknowledgements}
We thank the anonymous reviewer for their careful study and constructive comments. We gratefully acknowledge access to computational resources through the MIT Engaging cluster at the Massachusetts Green High-Performance Computing Center (MGHPCC) facility.

\appendix
\label{sec: Appendix}
\section{Kelvin-Helmholtz Contraction}
\label{sec: Kelvin-Helmholtz Contraction}
The interior structure of a star evolves considerably during the PMS phase. Here we adopt the model from \cite{Batygin2013ApJ} and summarize its key components. For a polytropic star with index $n$ and effective temperature $T_{\mathrm{eff}}$, we can assume that it radiates away energy, leading to the gravitational contraction of the star. To first order, we have
\begin{equation}
    \label{eq: stellar_E}
    \frac{dE}{dt} = -L_{\star} = -4\pi R_{\star}^{2}\sigma T_{\mathrm{eff}}^{4} = \frac{3}{10-2n}\frac{GM_{\star}^{2}}{R_{\star}^{2}}\frac{dR_{\star}}{dt},
\end{equation}
where $E$ is the energy of the star, and $L_{\star}$, $R_{\star}$, and $M_{\star}$ correspond to the stellar luminosity, radius, and mass, respectively.

Choosing a $n=3/2$ polytrope enables us to solve equation \ref{eq: stellar_E} assuming that the temperature $T_{\mathrm{eff}}$ is a constant, since this is valid only for convective stars. We also ignore any significant changes to the interior of the star and solve this differential equation to obtain the solution for the stellar radius,
\begin{equation}
    \label{eq: Rstar}
    R_{\star} = R_{\star 0}\left(1+\frac{t}{\tau_{R_{\star}}}\right)^{-1/3},
\end{equation}
where $R_{\star 0}$ is the initial stellar radius and $\tau_{R_{\star}}$ is the gravitational contraction timescale or the Kelvin-Helmholtz time. This can be calculated as
\begin{equation}
    \label{eq: tau_R}
    \tau_{R_{\star}} = \frac{3GM_{\star}^{2}}{28\pi\sigma T_{\mathrm{eff}}^{4}R_{\star 0}^{3}}.
\end{equation}
Setting $T_{\mathrm{eff}} = 4270 \ \mathrm{K}$ for a solar mass star with a radius larger than the Sun ($R_{\star 0} \gtrsim 4 \ R_{\odot}$), enables equation \ref{eq: Rstar} to approximate the evolution of a solar-type PMS star \citep{Siess2000A&A}. For typical values, $\tau_{R_{\star}} \approx 1 \ \mathrm{Myr}$.

\section{Angular Momentum Transport}
\label{sec: Angular Momentum Transport}
Gravitational contraction in isolation would lead to rotational spin-up, which can be solved analytically by solving
\begin{equation}
    \label{eq: grav}
    \left[\frac{dJ}{dt}\right]_{\mathrm{grav}} = 2I_{\star}M_{\star}R_{\star}\omega_{\star}\frac{dR_{\star}}{dt}+I_{\star}M_{\star}R_{\star}^{2}\frac{d\omega_{\star}}{dt} = 0,
\end{equation}
where $I_{\star}$ is the dimensionless moment of inertia of the host star and $\omega_{\star}$ is the stellar spin rate. Although the star accretes mass from the disk, the change in the star's mass is negligible, so we have assumed the stellar mass $M_{\star}$ to be constant while deriving and solving these equations. Solving equation \ref{eq: grav} using equation \ref{eq: Rstar} admits the solution
\begin{equation}
    \label{eq: grav_omega}
    \omega_{\star} = \omega_{\star 0}\left(1+\frac{t}{\tau_{R_{\star}}}\right)^{2/3}.
\end{equation}

The stellar magnetic field $\vec{B}_{\star}$ has dipole $B_{\mathrm{dip}}$ and octupole $B_{\mathrm{oct}}$ components \citep{Gregory2010RPPh, Adams2012ApJ}. When we include the terms of accretion and magnetic braking, equation \ref{eq: grav} becomes   
\begin{equation}
    \label{eq: dJ_dt}
    \left[\frac{dJ}{dt}\right]_{\mathrm{grav}} = 2I_{\star}M_{\star}R_{\star}\omega_{\star}\frac{dR_{\star}}{dt}+I_{\star}M_{\star}R_{\star}^{2}\frac{d\omega_{\star}}{dt} = \left[\frac{dJ}{dt}\right]_{\mathrm{accr}}+\left[\frac{dJ}{dt}\right]_{\mathrm{mag}}
\end{equation}
where the accretion term $[dJ/dt]_{\mathrm{accr}}$ and its corresponding timescale $\tau_{\mathrm{accr}}$ are given by \citep{Batygin2013ApJ},
\begin{equation}
    \left[\frac{dJ}{dt}\right]_{\mathrm{accr}} = \frac{\omega_{\star}}{\left(1+\Gamma\right)}\frac{R_{\star}^{2}}{2}\frac{M_{d_{0}}}{\tau_{d}}\frac{1}{(1+t/\tau_{d})^{2}} \ ,
\end{equation}
\begin{equation}
    \label{eq: tau_accr}
    \tau_{\mathrm{accr}} \approx 2I_{\star}(1+\Gamma)\tau_{d}\frac{M_{\star}}{M_{d_{0}}},
\end{equation}
where $\Gamma = B_{\mathrm{oct}}/B_{\mathrm{dip}}$ is the dimensionless ratio between the octopole and dipole components of the stellar magnetic field. If the star had no magnetic field, then we could solve equation \ref{eq: dJ_dt} analytically as there would be no contribution from the magnetic braking term (i.e. $[dJ/dt]_{\mathrm{mag}} = 0$). The solution for the stellar rotation rate would be
\begin{equation}
    \label{eq: accr_omega}
    \omega_{\star} = \omega_{\star 0}\left(1+\frac{t}{\tau_{R_{\star}}}\right)^{2/3}\exp{\left(\frac{1}{2}\frac{1}{(1+\Gamma)}\frac{M_{d_{0}}}{M_{\star}}\frac{t}{t+\tau_{d}}\right)}.
\end{equation}
For typical values for PMS stars and their corresponding accretion disks, the exponential factor in equation \ref{eq: accr_omega} is only a small correction to the solution given by equation \ref{eq: grav_omega}.

Before diving into the details of magnetic braking, we define three possible regimes for the star-disk magnetic interaction. The inner and outer edges of the magnetically connected region of the disk (in which closed magnetic field lines cross through the disk) are denoted by $\hat{a}_{\mathrm{in}}$ and $\hat{a}_{\mathrm{out}}$, respectively. The disk's truncation radius $a_{\mathrm{in}}$ is the inner edge where the star's magnetic pressure matches the ram pressure of the accreting material. The corotation radius $a_{\mathrm{co}} = \left(GM_{\star}/\omega_{\star}^{2}\right)^{1/3}$ represents the location where the centrifugal force owing to the disk's rotation cancels the gravitational pull of the host star. We also define $\mathcal{M} = |\vec{B}_{\star}(R_{\star},\pi/2)|R_{\star 0}^{3}/(1+\Gamma)$, the magnetic moment due to the equatorial field strength of the dipole magnetic field of the star. The stellar magnetic moment involves higher-order terms, but we can ignore them as they decay rapidly with the stellar radius. Finally, $\beta$ is the dimensionless diffusivity parameter \citep{Matt2005MNRAS} that parametrizes the magnetic coupling between the star and the disk; $\beta \gg 1$ corresponds to weak coupling and $\beta \ll 1$ to strong coupling.

Using these physical quantities, we can define the rate of angular momentum transport due to magnetic braking $[dJ/dt]_{\mathrm{mag}}$ and its corresponding timescale $\tau_{\mathrm{mag}}$ \citep{Batygin2013ApJ} under the no magnetic braking ($a_{\mathrm{in}} > \hat{a}_{\mathrm{out}}$), slow magnetic braking ($a_{\mathrm{in}} < \hat{a}_{\mathrm{in}}$) and disk-locked condition ($a_{\mathrm{in}} = a_{\mathrm{co}}$) \citep{Mohanty2008ApJ}:
\begin{equation}
\label{eq: mag}
\left[\frac{dJ}{dt}\right]_{\mathrm{mag}} =
    \begin{cases}
    0 & (a_{\mathrm{in}} > \hat{a}_{\mathrm{out}})\\[10pt]
    -\scalebox{1}{$\displaystyle\frac{4\pi}{3\mu_{0}}\frac{\beta\mathcal{M}^{2}}{a_{\mathrm{co}}^{3}(1+\beta)^{2}}$} & (a_{\mathrm{in}} = a_{\mathrm{co}})\\[20pt]
    -\scalebox{1}{$\displaystyle\frac{16\pi}{3\mu_{0}}\frac{\beta^{2}\mathcal{M}^{2}}{a_{\mathrm{co}}^{3}(1-\beta^{2})^{2}}$} & (a_{\mathrm{in}} < \hat{a}_{\mathrm{in}})
    \end{cases}
\end{equation}

\begin{equation}
\label{eq: tau_mag}
\tau_{\mathrm{mag}} \approx
    \begin{cases}
    \infty & (a_{\mathrm{in}} > \hat{a}_{\mathrm{out}})\\[10pt]
    \scalebox{1}{$\displaystyle\frac{3I_{\star}}{4\pi\omega_{\star 0}}\frac{(1+\beta)^{2}}{\beta}\frac{GM_{\star}^{2}\mu_{0}(1+\Gamma)^{2}}{\lvert\vec{B}_{\star}(R_{\star},\pi/2)\rvert^{2}R_{\star 0}^{4}}$} & (a_{\mathrm{in}} = a_{\mathrm{co}})\\[20pt]
    \scalebox{1}{$\displaystyle\frac{3I_{\star}}{16\pi\omega_{\star 0}}\frac{(1-\beta^{2})^{2}}{\beta^{2}}\frac{GM_{\star}^{2}\mu_{0}(1+\Gamma)^{2}}{\lvert\vec{B}_{\star}(R_{\star},\pi/2)\rvert^{2}R_{\star 0}^{4}}$} & (a_{\mathrm{in}} < \hat{a}_{\mathrm{in}})
    \end{cases}
\end{equation}
The $a_{\mathrm{in}} > \hat{a}_{\mathrm{out}}$ regime never undergoes resonance crossing since the radius contraction competes with the associated spin-up due to the lack of magnetic braking. However, in the presence of magnetic braking, the stellar spin-down tends to dominate in the later stages of the evolution, thus facilitating resonance crossing (Section \ref{sec: Resonance}). Although this holds for the $a_{\mathrm{in}} < \hat{a}_{\mathrm{in}}$ regime just as it does for the disk-locked case, the latter tends to occur much earlier as it maximizes magnetic braking, while the former might take longer than the lifetime of a typical accretion disk. 

\begin{figure}[ht!]
\centering
    \includegraphics[width=0.48\textwidth]{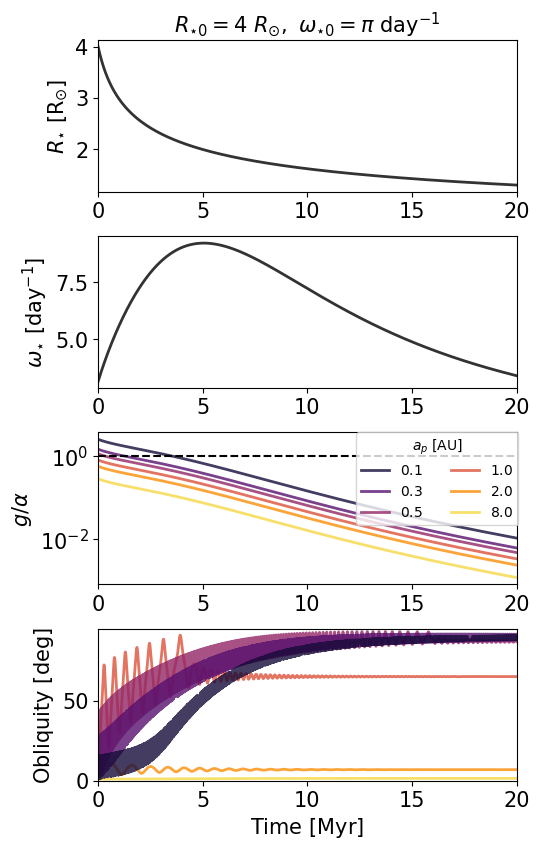}
    
    \caption{Planetary obliquity evolution from the secular model of the Hamiltonian in equation \ref{eq: H}, just like Figure \ref{fig: Obliquity_evolution_a}, but we instead assume magnetic braking without the disk-locked condition.}
    \label{fig: Obliquity_evolution_a_nolocked}
\end{figure}

To compare the $a_{\mathrm{in}} < \hat{a}_{\mathrm{in}}$ case with the $a_{\mathrm{in}} = a_{\mathrm{co}}$ case, we integrate the secular Hamiltonian (equation \ref{eq: H}) assuming $a_{\mathrm{in}} < \hat{a}_{\mathrm{in}}$ using the parameters in Table \ref{table: Model Parameters} and present its results in Figure \ref{fig: Obliquity_evolution_a_nolocked}. Figures \ref{fig: Obliquity_evolution_a} \& \ref{fig: Obliquity_evolution_a_nolocked} are not much different in terms of the overall obliquity excitation at each semi-major axis except for the borderline case at $a_{p} = 1 \ \mathrm{AU}$, which shows a larger obliquity excitation in the slower magnetic braking regime. However, there is a noticeable difference between the relevant timescales over which these resonance behaviors occur, with Figure \ref{fig: Obliquity_evolution_a} (disk-locked condition) experiencing adiabatic capture much earlier compared to Figure \ref{fig: Obliquity_evolution_a_nolocked} (slower magnetic braking). Using equation \ref{eq: tau_mag}, the typical timescale for the $a_{\mathrm{in}} = a_{\mathrm{co}}$ regime was estimated to be about $3 \ \mathrm{Myr}$ while that of the slower magnetic braking was around $100 \ \mathrm{Myr}$ \citep{Batygin2013ApJ}. The accretion disks of most PMS stars last $<10 \ \mathrm{Myr}$ \citep{Vlasblom2023A&A}. Therefore, in the context of this paper, the disk-locked condition is more important. While accretion also leads to spin-up, $\tau_{\mathrm{accr}}$ is about $40 \ \mathrm{Myr}$, which is longer than the disk lifetime, and hence has a negligible effect on the stellar rotation rate.

The solution for $\omega_{\star}$ cannot be obtained analytically once we include the non-trivial term in equation \ref{eq: mag}. We thus solve equation \ref{eq: dJ_dt} numerically and plot the stellar rotation period $P_{\star} = 2\pi/\omega_{\star}$ in Figure \ref{fig: P_rot} for three different initial conditions. For this calculation, we used a combination of observed and fiducial values for PMS stellar evolution for a solar-type star. $\Gamma = 1$ and $\lvert\vec{B}_{\star}(R_{\star},\pi/2)\rvert = 0.15 \ \mathrm{Tesla}$ were obtained from the observations of young T-Tauri stars \citep{Gregory2016arXiv}, while the other parameters ($M_{\star} = 1 \ M_{\odot}$, $R_{\star 0} = 4 \ R_{\odot}$, $M_{d_{0}} = 0.01 \ M_{\odot}$, $\tau_{d} = 0.5 \ \mathrm{Myr}$ and $\beta = 0.01$) are used as the fiducial values for the system. The dimensionless moment of inertia $I_{\star} = 0.2$ and the stellar love number $k_{2_{\star}} = 0.28$ for a polytrope $n=3/2$ were calculated using standard polytropic stellar models \citep{Chandrasekhar1939isss, Batygin2013ApJ}.

\section{N-body model for spin and orbit evolution}
\label{sec: Model for spin and orbit evolution}
In Section \ref{sec: N-body}, we perform $N$-body simulations to model more complex dynamics including stellar oblateness, disk-induced precession, multiple planets, and tides. Here we provide details of this code, which was originally developed by \cite{Millholland2019ApJ} and \cite{Spalding2020AJ}. The code works with direct accelerations in the framework of \cite{Mardling2002ApJ}, and it evolves the motion of each body by calculating all of the instantaneous accelerations that are applied to it. The dominant accelerations are those from Newtonian gravity, but other effects create perturbations and will be outlined here.

We calculate the acceleration on a planet $j$ due to the oblate star's ($\star$) quadrupole potential through
\begin{equation}
\label{eq: star_on_planet}
\begin{split}
    \vec{a}_{Q,\star j} &= \frac{k_{2_{\star}}}{2}\frac{R_{\star}^{5}}{r^{4}}\left(1+\frac{M_{p_{j}}}{M_{\star}}\right) \\
    & \left[\left(5\left(\vec{\omega}_{\star}\cdot\hat{r}\right)^{2}-{\omega}_{\star}^{2}-12\frac{GM_{p_{j}}}{r^{3}}\right)\hat{r}-2\left(\vec{\omega}_{\star}\cdot\hat{r}\right)\vec{\omega}_{\star}\right],
\end{split}
\end{equation}
where some of these variables were defined in the previous sections and where $M_{p_{j}}$ is the mass of the planet $j$ and $\vec{r}$ is the relative position vector from the star to the planet $j$. The stellar spin vector and its magnitude are $\vec{\omega}_{\star}$ and $\omega_{\star}$, respectively. This acceleration produces the planet's nodal recession due to stellar oblateness.

In addition, the tides raised by the star on a planet $j$ result in accelerations given by
\begin{equation}
    \vec{a}_{T,j} = \frac{-3n_{p_{j}}k_{2_{p_{j}}}R_{p_{j}}^{5}}{Q_{n_{p_{j}}}}\frac{M_{\star}}{M_{p_{j}}}\frac{a_{p_{j}}^{3}}{r^{8}}\left[3\left(\hat{r}\cdot\dot{\vec{r}}\right)\hat{r}+\left(\hat{r}\times\dot{\vec{r}}-r\vec{\omega}_{p_{j}}\right)\times\hat{r}\right]  
\end{equation}
where $n_{p_{j}}, k_{2_{p_{j}}}, R_{p_{j}}$ and $a_{p_{j}}$ correspond to the mean motion, Love number, radius, and the semi-major axis of the planet $j$. The spin vector and its magnitude of planet $j$ are $\vec{\omega}_{p_{j}}$ and $\omega_{p_{j}}$, respectively. $Q_{n_{p_{j}}}$ is the tidal quality factor of planet $j$ and $\dot{\vec{r}}$ is the time derivative of the relative position vector. The tidal quality factor is a dimensionless quantity that measures how efficiently a planet dissipates energy due to the tidal force from the host star. The tides raised by the planets on the star can be ignored as they do not play a significant role in the system's dynamics. 

Using the dimensionalized moment of inertia $I_{j}$ of planet $j$, we can calculate the spin evolution of the planet with the angular acceleration $\dot{\vec{\omega}}_{j}$ of the planet's spin axis,  
\begin{equation}
    I_{j}\dot{\vec{\omega}}_{j} = -\frac{M_{\star}M_{p_{j}}}{M_{\star}+M_{p_{j}}}\vec{r}\times\left(\vec{a}_{Q,j\star}+\vec{a}_{T,j}\right)
\end{equation}
where $a_{Q,j\star}$ is the acceleration on the star due to the quadrupolar gravitational moment of the planet $j$. It looks very similar to equation \ref{eq: star_on_planet} but with some variable subscripts switched between the star and the planet,
\begin{equation}
\begin{split}
    \vec{a}_{Q,j\star} &= \frac{k_{2_{p_{j}}}}{2}\frac{R_{p_{j}}^{5}}{r^{4}}\left(1+\frac{M_{\star}}{M_{p_{j}}}\right) \\
    & \left[\left(5\left(\vec{\omega}_{p_{j}}\cdot\hat{r}\right)^{2}-{\omega}_{p_{j}}^{2}-12\frac{GM_{\star}}{r^{3}}\right)\hat{r}-2\left(\vec{\omega}_{p_{j}}\cdot\hat{r}\right)\vec{\omega}_{p_{j}}\right].
\end{split}
\end{equation}

Finally, we will outline the acceleration components due to the gravitational potential of an infinite Mestel disk that we use in a subset of our $N$-body simulations. The surface density of the disk as a function of distance and time is
\begin{equation}
    \Sigma(r,t) = \Sigma_{0}(t)\left(\frac{r}{r_{0}}\right)^{-1},
\end{equation}
where the scaling factor $\Sigma_{0}(t)$ is parametrized to follow the evolution of the disk mass given by equation \ref{eq: disk_mass},
\begin{equation}
    \Sigma_{0}(t) = \frac{\Sigma_{0,0}}{1+t/\tau_{d}}.
\end{equation}
We next define the gravitational potential, which can be used to derive the acceleration due to the infinitely flat and extended protoplanetary disk \citep{Thommes2008ApJ, Spalding2020AJ}. The potential is
\begin{equation}
    \phi_{d} = 2\pi G\Sigma_{0}(t)r_{0}\ln{\left(|z|+\sqrt{z^{2}+r^{2}}\right)}.
\end{equation}
A disk potential of the form given above generates accelerations in the radial ($\hat{r}$) and vertical ($\hat{z}$) directions given by
\begin{equation}
\begin{split}
    a_{r} &= -\frac{\pi^{2}G\Sigma_{0}(t)r_{0}}{r}\left(1-\frac{|z|}{\sqrt{r^{2}+z^{2}}}\right) \\
    a_{z} &= -\pi^{2}G\Sigma_{0}(t)r_{0}\frac{\mathrm{sgn}(z)}{\sqrt{r^{2}+z^{2}}}.
\end{split}
\end{equation}
\bibliographystyle{aasjournalv7}
\bibliography{main}
\end{document}